\documentstyle[12pt,equations,psfig,cite]{article}
\def\bold#1{\setbox0=\hbox{$#1$}%
     \kern-.025em\copy0\kern-\wd0
     \kern.05em\copy0\kern-\wd0
     \kern-.025em\raise.0433em\box0 }

\def\slash#1{\setbox0=\hbox{$#1$}#1\hskip-\wd0\dimen0=5pt\advance
       \dimen0 by-\ht0\advance\dimen0 by\dp0\lower0.5\dimen0\hbox
         to\wd0{\hss\sl/\/\hss}}

\newcommand{\smallz}{{\scriptscriptstyle Z}} 
\newcommand{\smallw}{{\scriptscriptstyle W}} 
\newcommand{\smallh}{{\scriptscriptstyle H}} %
\newcommand{\smallr}{{\scriptscriptstyle R}} %
\newcommand{\eps}{\epsilon}

\newcommand{\acur}{ {\hat \alpha} }
\newcommand{\mz}{M_\smallz}
\newcommand{\mw}{M_\smallw}
\newcommand{\mh}{M_\smallh}
\newcommand{\mt}{m_t}

\newcommand{\sineff}{\mbox{$\sin^2 \theta^{{lept}}_{eff}$} }
 
\newcommand{\scur}{\mbox{$\hat{s}^2$}}
\newcommand{\sic}{\mbox{$\hat{s}$}}
\newcommand{\gc}{\mbox{$\hat{g}$}}
\newcommand{\sincur}{\mbox{$\sin^{2}\!\hat{\theta}_{\scriptscriptstyle W} 
                           (\mz^2)$}}
\newcommand{\ccur}{\mbox{$\hat{c}^2$}}

\newcommand{\rhoc}{\mbox{$ \hat{\rho}$}}
\newcommand{\dr}{\mbox{$ \Delta r$}}

\newcommand{\drcar}{\mbox{$\Delta \hat{r}$}}

\newcommand{\aww}{\mbox{$ A_{\smallw \smallw} $}}
\newcommand{\gmu}{\mbox{$ G_\mu $}}

\newcommand{\amtd}{\mbox{$ O(\alpha^2 m_t^2 /\mw^2) $}}
\newcommand{\amtq}{\mbox{$ O(\alpha^2 m_t^4 /\mw^4) $}}

\newcommand{\ew}{electroweak}
\newcommand{\msbar}{\overline{\rm MS}}
\newcommand{\ikp}{\int \frac{d^n k\, d^n p}{(2 \pi)^{2n}\, \mu^{-4\eps}}}
\newcommand{\equ}[1]{Eq.~(\ref{#1})}
\newcommand{\eqs}[1]{Eqs.~(\ref{#1})}

\newcommand{\be}{\begin{equation}}
\newcommand{\ee}{\end{equation}}
\newcommand{\een}{\end{subequations}}
\newcommand{\ben}{\begin{subequations}}
\newcommand{\beq}{\begin{eqalignno}}
\newcommand{\eeq}{\end{eqalignno}}
\newcommand{\bea}{\begin{eqnarray}}
\newcommand{\eea}{\end{eqnarray}}

\renewcommand{\thefootnote}{\fnsymbol{footnote} }
\begin{document}

\begin{titlepage}

\begin{flushright}
        \small
        MPI-PhT-96-77\\
        August 1996
\end{flushright}
\vspace{1.2cm} 
\begin{center}
{\Large\bf 
Precision tests of the Standard Model   \\
\vspace{.1cm}
           and higher order effects:} \\ 
\vspace{.2cm}
{\Large\bf  the example of the effective mixing angle }\\
\vspace{2.5cm}
{\sc  Paolo Gambino}\\
\vspace{.4cm}
\vspace{.3cm}
{\em Max Planck Institut f\"ur Physik, Werner Heisenberg Institut,\\
 F\"ohringer Ring 6, D80805 M\"unchen, Germany}
\end{center}
\vspace{3cm} 
\begin{center}
{\bf Abstract}
\end{center}
\vspace{0.2cm}
The radiative corrections involved in the precise determination of
the \ew\ mixing angle measured at the $Z^0$ peak
are reviewed in detail,
with particular emphasis on  the new
calculation of 
two-loop heavy top effects.
This example serves as a brief
pedagogical introduction to the problems and techniques 
of higher order \ew\ calculations.
\vspace{0.6cm}
\begin{center}{\em Lecture given at the 
XXXVI Cracow School of Theoretical Physics\\
Zakopane, Poland,  June 1996}
\end{center}
\vfill\nopagebreak

\end{titlepage}
\setcounter{footnote}{0}
\renewcommand{\thefootnote}{\arabic{footnote}}

The experimental verification of the Standard Model (SM) has 
reached in the last
decade a high degree of sophistication. On the experimental side we have now a 
large amount of very precise data collected at high energy, and several 
 observables are known at the permille level. On the theoretical side, 
a major effort has been undertaken in order to match the experimental 
accuracy. The one-loop corrections to all \ew\ observables are by now 
very well established \cite{YB},
and    two and three-loop effects have been investigated in many cases. 
Because of the number of different mass scales involved,   
multi loop calculations in the SM are highly non-trivial, but 
in the last few years a surge of activity in this field has made 
the investigation of some higher order effects  possible \cite{kniehl}.

Although the assessment of the theoretical error 
is a very subjective matter, we can generally distinguish between two 
different kind of uncertainties: {\em parametric}, i.e. due to the uncertainty
of the input parameters, and {\em intrinsic}, that is 
inherent to the truncation
of the perturbative expansion, or to the appearance of non-perturbative 
effects.
An important example of parametric uncertainty is the one due to the 
hadronic contribution
to the running of the electro-magnetic coupling from low energy
to the $Z^0$ resonance.
As  long distance dynamics is involved in this case,  one has to resort to 
the experimental data for $e^+ e^- \to hadrons$. Through the use of dispersion 
relations, $\alpha(\mz)$ is then determined with an accuracy of about
 $7\times 10^{-4}$ \cite{jeg}.
This sizable uncertainty is the main source of theoretical error for the 
$Z^0$-peak observables (except for low-angle Bhabha scattering).
On the other hand, scheme dependence, scale dependence,  
and explicit evaluation of higher 
order corrections can all  give information on the intrinsic error, hence the 
crucial importance of higher order corrections for the tests of the SM.

In this  lecture I will use a relatively simple example to illustrate
some of the problems and techniques involved in the evaluation
of the higher order effects relevant for the precision tests of the 
SM.
The case of  the effective Weinberg angle,
$\sineff$, measured at LEP and SLC, has  both the 
relevance and the simplicity to make a detailed and self-contained 
discussion possible. This observable is now known with excellent accuracy
(the present world average is $\sineff=0.23165\pm0.00024$ \cite{war}),
and its sensitivity to the Higgs mass is very high in comparison to other
quantities, giving it a predominant role in the present analyses.

After introducing to the basics of the renormalization of the SM,
I will  review the one-loop \ew\ corrections to $\sineff$.
 I will then discuss some potentially large higher order effects:
QCD corrections to hadronic loops, 
large mass two-loop effects (top quark and heavy Higgs case), etc.
 In the case of the two-loop heavy 
top corrections, I will describe in detail a new calculation of the 
\amtd\ contributions, and illustrate explicitly the heavy mass expansion 
method.

In the SM Lagrangian, besides the masses of the fermions and of the 
Higgs boson, and the CKM mixing parameters, there are only three independent
parameters, which we can choose to be, for example,
\be
g \ \ \ \ \ g' \ \ \ \ v, \hspace{1.5cm} {\rm or}\hspace{1.5cm}
 g \ \ \ \ \ \mw \ \ \ \  \mz,
\ee
where $g$ and $g'$ are  the $SU(2)_L$ and $U(1)_Y$ couplings, and
$v$ is the vacuum expectation value of the Higgs field. 
On the other hand, there are presently three very obvious experimental
inputs that we can use 
to determine these parameters: $\alpha$, the fine structure constant,
\gmu, the Fermi constant, and $\mz$; they are respectively known within
$4\times 10^{-9}$, $8\times 10^{-6}$, $2\times 10^{-5}$.
What effectively enters the physics of the weak interactions is generally
$\alpha(\mz)$, the running electro-magnetic coupling at the $Z^0$ scale. 
For the reasons mentioned above, this is known only slightly better 
than the permille level. 

A  number of natural  relations  link the couplings and the masses
of the gauge bosons occurring in the  bare SM  Lagrangian.
Although initially defined
in terms of the gauge couplings, after spontaneous symmetry breaking
the \ew\ mixing angle $\theta_\smallw^0$ relates both bare 
masses and couplings among themselves:
\be
\tan\theta_\smallw^0 = \frac{g_0'}{g_0}, \ \ \ e_0=g_0 \sin\theta_\smallw^0,
\ \ \   \mw^0 = \cos\theta_\smallw^0 \mz^0.
\label{relations}
\ee
At any order in perturbation theory we can relate the parameters of the 
Lagrangian to the measured inputs, calculate them, and then make predictions
for any observable. For instance, using 
$s\equiv \sin\theta_\smallw$ and $c\equiv \cos\theta_\smallw$,
most relevant among such relations are
\ben
\beq
\ \ s^2= 
\frac{\pi\alpha}{\sqrt{2}G_\mu \mw^2} \hspace{1.5cm}\rightarrow
 \hspace{1.3cm}
\ s^2_\smallr= 
\frac{\pi\alpha}{\sqrt{2}G_\mu \mw^2} \frac1{\left(1-
\dr_\smallw^\smallr\right)}
\eeq
\beq
\mw^2= c^2 \ \mz^2
\hspace{1.3cm} \rightarrow  \hspace{1.3cm}
\mw^2= c^2_\smallr \ \mz^2 \ \rho_\smallr.
\label{drho}
\eeq
\label{radcorr}
\een
Here the l.h.s. corresponds to a tree level description, 
while on the r.h.s. I have shown 
that, in a given renormalization scheme "R", the quantum 
effects can be incorporated through the radiative corrections
$\dr_\smallw^\smallr$ and $\rho_\smallr$. 
They are functions of $s_\smallr$, $\mw$, $\mt,\ \mh$, etc.,
and clearly depend on the renormalization scheme.
In the following I will adopt  $\msbar$ renormalized couplings at the $Z^0$
scale,
so that the mixing angle is defined by 
\be
\scur\equiv
\sin^2\hat{\theta}_\smallw (\mz)_{\msbar}\equiv
\frac{\hat{\alpha}(\mz)_{\msbar}}{\hat{\alpha}_2(\mz)_{\msbar}}
\label{msbarsine}
\ee
($\hat{\alpha}$ and $\hat{\alpha}_2$ are the $\msbar$ U(1)$_{e.m.}$ 
and SU(2) running couplings),
while the vector boson masses are defined as the physical masses,
as in \equ{radcorr} \cite{msbar}. 
This choice has some advantages which will be clear later.
In this framework\footnote{Alternative approaches are reviewed
in \cite{hollik}; the on-shell scheme was originally proposed 
in \cite{ms80}.}, $\Delta\rhoc\equiv 1- \rhoc^{-1}$ and $\drcar_\smallw$
are the radiative corrections entering the l.h.s. of \equ{radcorr}.
By solving these two equations simultaneously, 
the mass of the $W$ boson and the $\msbar$ mixing angle can be determined.

At LEP and SLC $\sineff$ is measured  from the on-resonance asymmetries.
Leaving aside the photon mediated amplitudes and the 
QED radiative corrections,
which form a finite and gauge invariant set and are routinely subtracted
by the experimental groups, the left-right (LR) and forward-backward 
asymmetries depend only on the ratio of the vector and axial-vector couplings 
of the $Z^0$ to the leptons. In the simple case of the LR asymmetry measured
 at SLC, 
\be
A_{LR}= \frac{2 \,g_V^\ell \,g_A^\ell}{(g_V^\ell)^2 + (g_A^\ell)^2},
\ee
where $g_V^\ell$  and $ g_A^\ell$ are the vector and axial-vector couplings
of an on-shell $Z^0$ to the  leptons.

At the tree level, the amplitude for the decay of a $Z^0$ boson
on mass-shell is given by:
\centerline{
\psfig{figure=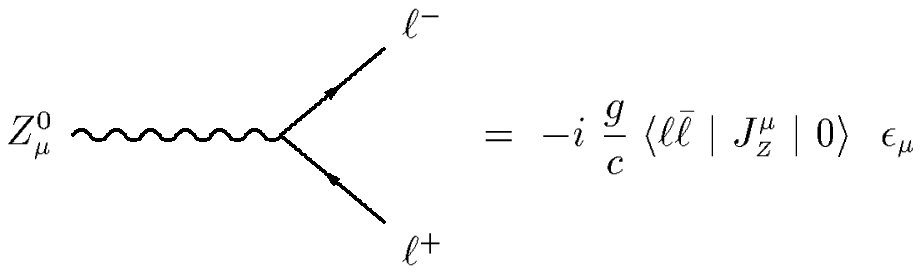,rheight=1.19in}  }
\label{tree}
where $\epsilon_\mu$ is the polarization vector of the $Z^0$, and 
$J_\smallz^\mu$ is the fermionic current coupled to the $Z^0$. The latter can 
also be written as 
\be
J_\smallz^\mu=
\frac1{2} \ J_3^\mu - s^2 \ J_\gamma^\mu,
\ee
with $J_\gamma^\mu= Q_q \ 
\bar{q}\gamma^\mu  q$, and $J_3^\mu= I_3^q \ \bar{q}\gamma^\mu a_-  q$\,
the $U(1)$ and $SU(2)_L$ currents respectively
($I_3^q=\pm1$ and
$a_\mp= \frac1{2} (1\mp \gamma_5)$). 
We then define $ 1- 4 \sineff\equiv  g_V^\ell/g_A^\ell $, and $A_{LR}$
directly measures $\sineff$.

\begin{figure}[h]
\centerline{
\psfig{figure=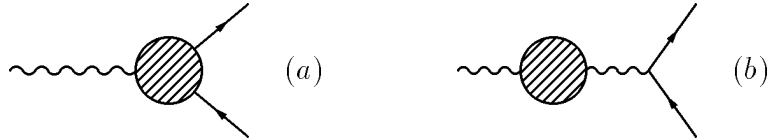,rheight=0.71in}  }
\caption{\sf One-loop Feynman diagrams contributing to $Z^0$ decay amplitude.} 
\label{oneloop}
\end{figure}

At the {\bf one-loop}
 level much of this simplicity is retained. We need to consider
only the \ew\ effects; the two  
topologies that contribute to the $Z$-decay amplitude at this order are 
depicted in Fig.\ref{oneloop}, 
where $(a)$ 
summarizes all vertex corrections and the wave function 
renormalization of the external fermions, and 
$(b)$ represents a self-energy insertion, that either mixes 
the $Z$ with the photon, or contributes to the wave function renormalization
 of the $Z^0$, in which case it is multiplied by a factor $1/2$.
Just because of their tensorial structure, the sum of these contributions
and of the tree level one
can be written in terms of {\em bare} quantities as
\be
- i\ \frac{g_0}{c_0} \ A \ \langle \bar{\ell} \ell \ | \ \frac1{2}
J_3^\mu - s_0^2 \ B \, J_\gamma^\mu \, | \  0\rangle,
\label{loopam}
\ee
where $A$ and $B$ are two divergent quantities, regulated
in $n$ dimensions. At this point we perform the 
renormalization of the couplings by a $\msbar$ subtraction at the $Z^0$
scale. In \equ{loopam} we
 replace $s_0$  by $\hat{s}(\mu) - \delta \hat{s}$, and $e_0= g_0 \ s_0$ by 
$\hat{e}(\mu) - \delta \hat{e}$, and set the scale  $\mu=\mz$.
Notice that $\delta \hat{s}$ and $\delta \hat{e}$ are pure poles in 
$\epsilon=  (4-n)/2$. Expanding everything up to $O(\hat{\alpha})$
we get rid of the divergences. The new amplitude is finite:
\be
- i\ \frac{\hat{g}}{\hat{c}} \ \sqrt{\bar{\rho}}\ 
 \langle \bar{\ell} \ell \ | \ \frac1{2}
J_3^\mu - \scur\  \hat{k}(\mz^2)\, J_\gamma^\mu \, | \ 0\rangle,
\label{loopam2}
\ee
and we can identify 
\be
\sineff = {\rm Re} \ \hat{k}(\mz^2) \ \scur(\mz).
\ee

In order to
 actually compute \sineff, we therefore need the form factor $\hat{k}(\mz^2)$,
which depends only on the $J^\mu_\gamma$
 component of the diagrams in Fig.\ref{oneloop}. This is given by the mixing
amplitude, $-i \gc \sic \,\langle\bar{\ell} \ell  | J_\gamma^\mu
| 0\rangle\  A_{\gamma Z}(\mz^2)/\mz^2$, 
and by the part of the vertex amplitude proportional to $ J_\gamma^\mu$,
$-i \gc/\hat{c}\ \langle\bar{\ell} \ell  | J_\gamma^\mu | 0\rangle\ \scur 
\, V_\gamma$. Hence
\be
\hat{k}(\mz^2)= 1- \frac{\hat{c}}{\sic}\, \frac{A_{\gamma Z}(\mz^2)}{\mz^2}
- V_\gamma - \frac{\delta\scur}{\scur}.
\label{dk}
\ee
From the ultraviolet divergent part (see for ex. the second paper in
 Ref.\cite{ms80})
of the vector boson mass counterterms
we calculate the counterterm $\delta\scur=-\delta\hat{c}^2=-[ \delta(
\mw^2/\mz^2)]_{UV}$,
\be
\frac{\delta\scur}{\scur}= - \frac{\gc^2}{16\pi^2}\left(
\frac{19}{6} + \frac{11}{3}\,\scur\right)\ \frac1{\epsilon}. 
\label{ds2}
\ee
In the t'Hooft-Feynman gauge the divergent part of the mixing self-energy 
is \cite{ms80}
\be 
\frac{A_{\gamma Z}(\mz^2)}{\mz^2}= \frac{\gc^2}{16\pi^2}
\,\frac{\sic}{\hat{c}}\,\frac1{\epsilon} \,\left[
\frac7{6} +\frac{17}{3}\scur + 2\frac{\mw^2}{\mz^2} -2\hat{c}^2+O(\epsilon)
\right].
\label{agz}
\ee
 Similarly, the divergence
of \ $V_\gamma$ \  is $2\, \gc^2\,\hat{c}^2/(16\pi^2)$ $1/\epsilon$.
Using Eqs. (\ref{dk}-\ref{agz}), and the l.h.s. of \equ{drho}, we see that
the form factor $\hat{k}(\mz^2)$ is indeed finite through $O(\acur)$. It is 
also gauge invariant and independent of the Higgs mass, as the Higgs boson 
does not couple to the photon, and the Higgs vertices are 
suppressed by the Yukawa couplings of the leptons. 
Explicit expressions of $\hat{k}(\mz^2)$ can be found in \cite{ds,gs}.

Because of the top loop contribution to $A_{\gamma Z}(\mz^2)$, however,
the form factor does depend on the top mass, albeit very 
weakly\footnote{For simplicity, $\scur$ is defined
here without  decoupling the top quark. That approach is considered in 
\cite{gs,FKS}.}:
\be
\hat{k}(\mz^2)_{top}= -\frac{\acur}{6\pi\scur} \left(1-\frac8{3}\scur\right)
\ln \frac{\mt^2}{\mz^2} \ + \, O({\mz^2\over\mt^2}).
\label{top}
\ee

Numerically, the real part of the form factor is very close to one 
\cite{gs}. For $\mt=175$GeV, $\mw=80.356$GeV, and $\scur=0.2316$,  
\be
\sineff = \scur(\mz^2) + 4 \times 10^{-5}
\label{sin1}
\ee
The unusual smallness of the one-loop \ew\ radiative corrections 
($\approx 2\times 10^{-4}$) is due to large cancellations between bosonic and
fermionic contributions, which are individually over 
one order of magnitude larger
\cite{gs}. The imaginary part of $\hat{k}(\mz^2)$, 
although comparatively large,
gives a  negligible contribution to the asymmetries \cite{gs}.

\equ{sin1} gives the  one-loop prediction for \sineff 
in the SM: once \scur\ is calculated
from the inputs via the l.h.s. of  \eqs{radcorr}, \sineff\ can be known with a
great accuracy. Of course, \scur\ depends very sensitively on
$\mh$ and $\mt$, through $ \drcar_\smallw$ and $\rhoc$, and has an 
 uncertainty from the value of $\alpha(\mz)$ of about 0.1\%. Moreover,
unknown higher order contributions  could add a significant 
uncertainty to its determination. In that respect, however, 
 the situation for \scur\ seems to be reasonably under control 
\cite{sirqcd,crad} after the 
calculation of the three-loop QCD \cite{chet} and of the  two-loop 
\amtd\ \cite{us,new} corrections.

Here I will concentrate on the analysis of higher order effects on 
\equ{sin1}. Given the estimated intrinsic accuracy on \scur,  we can aim 
to a precision of 
about $10^{-4}$ on the form factor  $\hat{k}(\mz^2)$.  
 A first candidate are the {\bf QCD corrections} to the hadronic loops.
Fortunately, the scale of the process at hand is such that  perturbative
 QCD can be reliably applied. As  quark loops appear only in the amplitude
 of Fig.\ref{oneloop}(b),
QCD   affects only the 
$\gamma-Z$ mixing as in the figure below, 
where 
the lowest order diagrams are displayed. 
Since the axial and the vector current do not mix even in presence of QCD,
only the vector current correlator contributes to $A_{\gamma Z}(\mz^2)$,
reducing the two-loop calculation to the old QED one \cite{Jost}.
\begin{figure}[h]
\centerline{
\psfig{figure=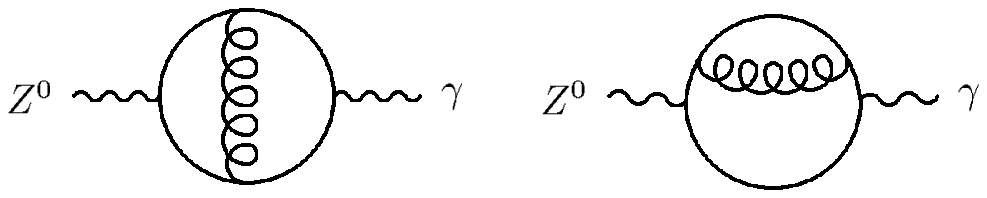,rheight=0.71in}  }
\end{figure}
However, two-loop QCD corrections to the general \ew\ vector boson and scalar
self-energies are known exactly \cite{abdel}
for any value of the quark masses and the external momentum, in terms
 of simple logs and dilogs. 
Even the three-loop massless quark contributions
and  the heavy top expansions
of the same correlators are now available \cite{chet,massless}.

Unlike the case of the $\rho$ parameter, however, QCD corrections are not
important for the form factor $\hat{k}(\mz^2)$. Using \cite{abdel},
we see that the light quark 
loops, neglecting all mass effects, get a correction
\be
\Delta \hat{k}(\mz^2)_{QCD}^{light}= \frac{\acur}{\pi \scur}
\left( \frac7{12} - \frac{11}{9} \scur\right)
\frac{\alpha_s(\mz)}{\pi} \left( \frac{55}{12} - 4 \zeta(3) + i \pi\right),
\ee 
whose real part
 amounts to about $-3\times 10^{-5}$. Using the expansions in the 
third paper of Ref.\cite{abdel}, we can  see that the top contribution
of \equ{top} can be adjusted to include QCD corrections by the replacement
$\log \mt^2/\mz^2\to \log\mt^2/\mz^2 \ (1 + \alpha_s(\mt)/\pi) - (15/4) \,
 \alpha_s(\mt)/\pi$. Numerically, these QCD corrections contribute about 
$7\times 10^{-5}$ to $\hat{k}(\mz^2)$.
It seems that  there is no reason to go beyond two-loop level in QCD.

\begin{figure}[h]
\centerline{
\psfig{figure=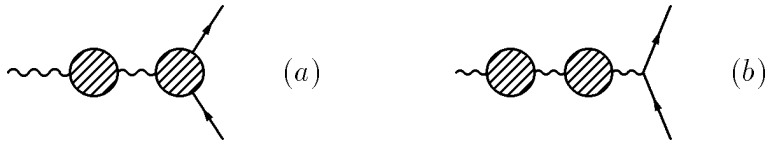,rheight=0.57in}  }
\caption{\sf Main  two-loop \ew\ topologies contributing to $k$. } 
\label{twoloop}
\end{figure}

A relatively large two-loop effect which can be easily accounted for
 is given by the
{\bf interference of the imaginary parts} of $A_{\gamma Z}(\mz^2)$ and of the 
photon self-energy (Fig.\ref{twoloop}(b)).
Indeed, light fermion loops induce large and additive 
imaginary parts in these two 
self-energies. We can resum the gauge-invariant fermionic part of the photon
self-energy  multiplying $A_{\gamma Z}(\mz^2)$ in \equ{dk} 
by $\mz^2/(\mz^2 - A_{\gamma\gamma}^{(f)}(\mz^2)|_{\msbar})$ \cite{ds},
where $A_{\gamma\gamma}^{(f)}(\mz^2)|_{\msbar}$ is the fermionic part of 
the photon
two-point function  renormalized in $\msbar$. The interference between 
the two imaginary parts (the contribution of the real parts is negligible) 
increases the value of $ \hat{k}(\mz^2)$ by 1.9$\times 10^{-4}$, a quite large 
two-loop effect. The inclusion of this kind of contributions is necessary.

Another potential source of large higher order effects are heavy particles;
it is well-known that they do not decouple in the SM, and that
the corrections
due to the heavy top are dominant in $\hat{\rho}$  at the one-loop level.
The study of the leading effects of this origin is therefore
very important, and is made possible by an expansion in powers of the 
heavy masses.
Let us start briefly considering  the possibility of {\bf heavy Higgs} effects.

At one loop, the  radiative corrections to amplitudes without Higgs bosons
among the external particles depend at most logarithmically on the Higgs
mass, in the limit in which this mass is heavy. At two-loop the corrections
can be at most quadratic in the Higgs mass. For the $\rho$ parameter
these quadratic terms
were computed long ago, and  more  recently a complete analysis for \ew\
observables has been performed \cite{veltbarb}, in  the case where the Yukawa 
couplings are neglected.  
The result is that these corrections are completely negligible.
Still,  the Higgs boson could
appear in the irreducible two-loop $\gamma-Z^0$
mixing in association with  its Yukawa coupling with the top.
 In this case, however,
the amplitude is also proportional to $\mt^2$, and is best  considered
in connection with a heavy top.

Finally, we consider the two-loop effects of a {\bf heavy top quark}.
In the case of the amplitude at hand, we have seen that, at one-loop, the top 
contributes  in a mild, logarithmic  way (see \equ{top}). At two loops, by 
analogy, we expect only \amtd\ effects, compared with
the \amtq\ appearing in $\Delta\hat{\rho}$ \cite{mt4}. 
Following the recent calculation
of the \amtd\ effects on \equ{radcorr} \cite{us}, which turn out to be
relatively important \cite{new}, 
the study of  these quadratic two-loop contributions
is necessary. The rest of this lecture is devoted to a detailed presentation
of such calculation.

At the two-loop level the diagrams contributing to the $Z^0$ decay amplitude
and proportional to $J_\gamma^\mu$ can be grouped into three categories: 
(i) products of one-loop $Z^0$ wave function renormalization by one-loop
vertices and self-energies (Fig.\ref{twoloop} with the first two wavy lines
representing $Z^0$'s);
(ii) products of one-loop $\gamma-Z^0$ mixing amplitudes by 
vertices and self-energies (Fig.\ref{twoloop} with the second wavy line 
representing a photon);
(iii) two-loop irreducible vertices and $\gamma-Z^0$ self-energy, displayed in 
Figs.\ref{oneloop}(a,b).
 
We are interested only in the two-loop contributions proportional to $\mt^2$.
It is  easy to realize that only the last category meets this condition:
we have to keep in mind that the one-loop vertices do not contain the top quark
and 
that the $\gamma-Z^0$ self-energy depends only logarithmically on $\mt$
(\equ{top}). We also  notice that the wave 
function renormalization factor of the $Z^0$   contains only $\log\mt$
in the heavy top limit.
This can be understood on dimensional grounds, as this factor 
 involves only the derivative of the $Z^0$ vacuum polarization
function w.r.t. the external momentum.  We therefore need to consider only the
subset of  
two-loop irreducible vertices and  $\gamma-Z^0$ self-energies which can be
 proportional to $\mt^2$.

In addition to the two-loop diagrams, we need to include the $\mt^2$ part
of the two-loop $\msbar$
counterterm of $\scur$, and renormalize the parameters appearing in 
the one-loop correction \equ{dk}. With our choice of $\msbar$ couplings and 
on-shell masses, when does renormalization of the one-loop amplitude 
introduce $\mt^2$ terms? From \equ{ds2} we see that $\delta\scur$ does not 
contain any such term, and the same is trivially true of $\delta\acur$.
As the one-loop amplitude is at most logarithmic in $\mt$, we conclude that 
the renormalization of the couplings does not contribute at the level of our
calculation. On the other hand, the mass renormalization 
of the top quark and of the vector 
bosons does  contribute $O(\mt^2)$ terms, because of the top dependence
of the on-shell mass counterterms. The Goldstone bosons and the ghost 
fields that appear
 in the one-loop loops also need mass renormalization.
In the  t'Hooft-Feynman gauge employed here, this is implemented
assigning them the corresponding vector boson on-shell mass counterterm.
In the case of the Goldstone bosons, however, a spurious quartic $\mt^4$
dependence must be cancelled by a tadpole contribution, as dictated by
Ward identities.

At \amtd, we can finally write the two-loop form factor as
\be 
\hat{k}^{(2,top)}(\mz^2)= 1- \frac{\hat{c}}{\sic}\, 
\frac{A_{\gamma Z}^{(2+c.t.)}(\mz^2)}{\mz^2}
- V_\gamma^{(2+c.t.)} - \frac{\delta^{(2)}\scur}{\scur} + \Delta_{extra},
\label{dk2}
\ee
where the term $\Delta_{extra}$ is  
generated in \equ{agz} by reexpressing
the ratio $2\mw^2/\mz^2$ in terms of $\ccur$, using the l.h.s. of \equ{drho}.
As $\Delta\rhoc$ is $O(\acur\mt^2/\mw^2)$, this introduces a crucial term
for the finiteness of our calculation. In $n$ dimensions it 
is necessary to keep $O(\epsilon)$ terms in $\Delta\rhoc$, so that a
finite part is generated: $\Delta_{extra}=
-N_c \,\gc^4/(16\pi^2)^2 \ \mt^2/\mz^2 \ 1/(2\epsilon)\, +$\, finite part,
where $N_c=3$
is the number of colors.
The two-loop counterterm $\delta^{(2)}\scur$ is a crucial component of
the calculation in \cite{us}, can be checked with \cite{macha}, and reads 
\be
\frac{\delta^{(2)}\scur}{\scur}=
-\frac{N_c\, \gc^4}{(16\pi^2)^2}\, \frac{\mt^2}{\ccur\mz^2}\,\frac1{\epsilon}
\left(\frac1{8} - \frac{13}{36} \,\scur\right).
\ee

\begin{figure}[h]
\centerline{
\psfig{figure=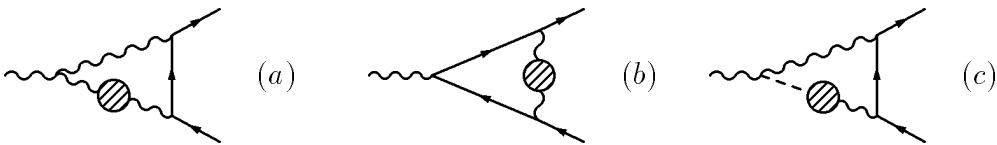,rheight=0.73in}  }
\caption{\sf Two-loop vertex diagrams containing the top.}
\label{vert2}
\end{figure}

We now begin the study of the irreducible diagrams with the
vertices. The presence of the top quark is only induced by loop insertions
on the internal vector boson propagators.    
The relevant diagrams are shown in  Fig.\ref{vert2}(a,b).
 Scalar loops are generally suppressed by
powers of the lepton mass, but at two-loop level the mixing $\phi^+ -
W^+$ takes place through the top in the diagram of Fig.\ref{vert2}(c).
For what concerns the diagrams of Fig.\ref{vert2}(a,b), 
we observe that the leading $\mt^2$ contribution is cancelled by the 
mass renormalization of the corresponding vector boson. 
As the overall divergence of such graphs is logarithmic, and the quadratic 
subdivergence of the top loop is the same as in the counterterm
$\delta\mw^2=\aww(\mw^2)\approx \aww(0)$, the use of on-shell masses in the
one-loop vertices removes all \amtd\ vertex contribution at two-loop level.
This shows a clear  advantage of the renormalization procedure adopted.
The case of Fig.\ref{vert2}(c) is obviously very different and requires
an explicit two-loop calculation, which I describe as an example
of the heavy mass expansion method.

A few kinematical simplifications
allow us to write the amplitude in Fig.\ref{vert2}(c) as
\be
I=-i\, \frac{\gc^5 \, \scur\, \mt^2}{2\,\hat{c}}
\,\langle\ell\bar{\ell}\, |J_3^\mu|\, 0\rangle \ikp \
\frac{k\cdot p}{k^2\, p^2\,  P_\smallw^2 \, PK_t \, PQ_\smallw},
\label{I}
\ee
where I used the abbreviations $P_\smallw= p^2-\mw^2$, $PK_t=(p-k)^2-\mt^2$,
and $PQ_\smallw=(p-q)^2-\mw^2$.
First we notice that $I$ is in fact a self-energy integral,
because all the dependence on the external lepton momenta has dropped out.
It can therefore be treated as a two-loop two-point function.

In general, a two-loop two-point function cannot be expressed in terms
of polylogarithms  or other known functions (of course in special cases,
like for QCD corrections considered above, the two-loop
integrals can be expressed in a compact way). The choice is then between
numerical evaluation, which can be now very efficient \cite{num},
and approximate analytic evaluation. In particular, we are now interested 
in a heavy mass expansion \cite{asym} that
enables us 
to  expand the integral in \equ{I} in powers of $\mt$, keeping only
the leading $\mt^2$ term. 
The simple 
method here illustrated is suggestive of the strategy adopted in \cite{us}.

The integral $I$ has a threshold at $q^2=\mw^2$,
and is therefore  not analytic at $q^2=\mz^2$.
 We cannot use a simple Taylor expansion
in the external momentum in the approximation $q^2\ll \mt^2$, but we have to 
resort to an asymptotic expansion. The identity 
\be
\frac1{(p-k)^2-\mt^2} \to \frac1{k^2-\mt^2} +
\frac{2\,p\cdot k - p^2}{(k^2-\mt^2)\left[(p-k)^2-\mt^2\right]}
\label{id}
\ee
helps us separate different regions of the domain of integration.
Using  \equ{id} 
 in $I$,  we obtain  a
disconnected two-loop integral (product of two one-loop integrals), plus a
two-loop integral with improved ultraviolet convergence in the $k$
integration and improved
infrared convergence in the $p$ integration.
Applying \equ{id} twice to $I$, we obtain, up to disconnected integrals
proportional to $k\cdot p$ that vanish by symmetry,
\be
\ikp \left[\frac{2(k\cdot p)^2}{k^2\, p^2\,  K_t^2 P_\smallw \, PQ_\smallw} -
 \frac{k\cdot p \ (2 \,k\cdot p - p^2)^2}{k^2\, K_t^2\, p^2\,
  P_\smallw^2 \, PK_t \, PQ_\smallw}\right]
\ee
with $K_t=k^2-\mt^2$.
The first integral can be easily
reduced to  the product of two well-known one-loop integrals, while the second
is $O(1/\mt^2)$, as can be checked counting the degrees of infrared and
ultraviolet divergence, and can be dropped.
Finally, we are able to write the vertex contribution proportional to 
$J_\gamma^\mu$ as
\be
V_\gamma^{(2+c.t.)}= - \frac{N_c\, \gc^4 \, \scur}{(16\pi^2)^2 } \,
\frac{\mt^2}{\mz^2}
\left[ \frac{C0[\mz^2,\mw,\mw]}{\eps} + {\rm finite } +
O\left(\frac{1}{\mt^2}\right) \right],
\ee
where $C0[q^2,m_1,m_2]$ is a convergent one-loop integral.
The method followed in this example can be used in a similar way in the 
case of the irreducible $\gamma-Z^0$ self-energies. Schematically, we can 
summarize the effect of the heavy top expansion as in the figure above,
\begin{figure}[h]
\centerline{
\psfig{figure=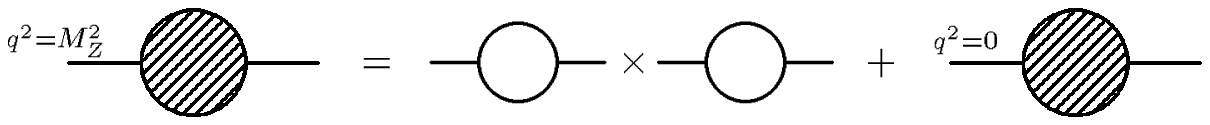,rheight=0.42in}  }
\end{figure}
where the one loop diagrams are evaluated exactly at  the appropriate momentum
$q^2=\mz^2$. The two-loop vacuum ($q^2=0$) integrals can always be reduced
to a closed analytic expression containing  dilogarithms following
the strategy proposed, for ex., in \cite{DT}.
For what concerns the counterterm pieces, the procedure consists simply 
of expanding all masses 
in the one-loop $A_{\gamma Z}(\mz^2)$ according to $m\to m-\delta m$,
 keeping
all $O(\eps)$ terms. The combination of irreducible and counterterm
pieces is
\be
A_{\gamma Z}^{(2+c.t.)}(\mz^2)=
\frac{N_c \, \gc^4\,\sic\,\mt^2}{(16\pi^2)^2 \hat{c}^3}
\left[\left(\scur\,\hat{c}^2 \, C0[\mz^2,\mw,\mw] -\frac3{8} +
\frac5{36}\scur\right)\,\frac1{\epsilon} + {\rm finite}\right]
\ee
Putting together the various pieces of \equ{dk2} the cancellation of 
divergences can be verified. 
In the notation of \cite{us},
the final result  for  ${\rm Re}\,\hat{k}^{(2, top)}(\mz^2)$,
in units 
$N_c (\hat{\alpha}/(4\pi \hat{s}^2))^2$ $ m_t^2/M_W^2$, reads
\begin{eqnarray}
&{\rm Re}\,&\Delta\hat{k}^{(2,top)}(\mz^2)=
{{-211 + 24\,{\it ht} + 462\,{\scur} - 64\,{\it ht}\,{\scur
}}\over {432}} + 
  \left( {3\over 8} - {{{\scur}}\over 3} \right) 
   B0[\mz^2,\mw^2,\mw^2] 
\nonumber\\&&
+  {{\left( ht-4 \right) \,{\sqrt{{\it ht}}}\,
      \left(  8\,{\scur}-3 \right) g(ht)}\over {108}} - 
  {{\left( 6 + 27\,{\it ht} - 10\,{{{\it ht}}^2} + {{{\it ht}}^3} \right) \,
      \left( 3 - 8\,{\scur} \right) \ln {\it ht}}\over 
    {108\left( {\it ht}-4 \right) }} 
\nonumber\\&&
- {\ccur\over6} \ln \ccur +
  \left( \frac5{36} + \frac7{18} \scur\right) \ln zt + 
  {{\left(  {\it ht}-1 \right) \left(  8\,{\scur} -3\right) 
     }\over 
    {18\left( 4 - {\it ht} \right) {\it ht}}} \, \phi ({ht\over 4}).
\end{eqnarray}
Numerically, this correction is negligible. It grows with $\mh$, reaching 
about 2.5$\times 10^{-5}$ for $\mh=700$GeV, well below our  precision
goal. However, it must be emphasized that this result is strictly
related to the renormalized parameters that have been chosen here,
and to the way the one-loop result has been expressed.
When all the two-loop effects considered here are included,
and using the same inputs, \equ{sin1} is modified into
\be
\sineff = \scur(\mz^2) + 1.0 \times 10^{-4}.
\ee

In summary, I have reviewed the radiative corrections entering the relation
between the effective leptonic sine and the $\msbar$ parameter $\sincur$.
The main higher order effects have been considered, and a new calculation
of the \amtd\ contributions has been illustrated in detail. All these 
corrections 
are very small and  we can conclude  that in the case at hand
higher order effects are unlikely to 
be a major source of uncertainty.

\vspace{4mm}
I am grateful to Marek Jezabek and to the organizers of the School 
for the kind invitation, the  hospitality, 
 and  the excellent organization. 
The results of the two-loop heavy top calculation have been obtained in 
collaboration with G. Degrassi.

\end{document}